\documentclass[aps,pr,twocolumn,superscriptaddress,noshowpacs,a4paper,10pt]{revtex4}
\pdfoutput=1
\usepackage{color}
\usepackage{dcolumn} 
\usepackage{amssymb}
\usepackage{amsmath}
\usepackage{graphicx}
\usepackage[pdftex]{hyperref}
\hypersetup{colorlinks=true,citecolor=blue,linkcolor=blue,urlcolor=blue}
\usepackage[all]{hypcap}

\definecolor{green}{rgb}{0,0.5,0}

\begin{document}
\title{Static friction boost in edge-driven incommensurate contacts}

\author{
  \firstname{Davide} \surname{Mandelli}}
  \affiliation{Department of Physical Chemistry, School of Chemistry,
  The Raymond and Beverly Sackler Faculty of Exact Sciences and The
  Sackler Center for Computational Molecular and Materials Science,
  Tel Aviv University, Tel Aviv 6997801, Israel}
\author{\firstname{Roberto} \surname{Guerra}}
  \affiliation{Center for Complexity and Biosystems, Department of Physics, University of Milan, 20133 Milan, Italy}
\author{\firstname{Wengen} \surname{Ouyang}}
  \affiliation{Department of Physical Chemistry, School of Chemistry,
  The Raymond and Beverly Sackler Faculty of Exact Sciences and The
  Sackler Center for Computational Molecular and Materials Science,
  Tel Aviv University, Tel Aviv 6997801, Israel}
\author{\firstname{Michael} \surname{Urbakh}}
  \affiliation{Department of Physical Chemistry, School of Chemistry,
  The Raymond and Beverly Sackler Faculty of Exact Sciences and The
  Sackler Center for Computational Molecular and Materials Science,
  Tel Aviv University, Tel Aviv 6997801, Israel}
\author{\firstname{Andrea} \surname{Vanossi}}
  \affiliation{CNR-IOM Democritos National Simulation Center, Via Bonomea 265, 34136 Trieste, Italy}
  \affiliation{International School for Advanced Studies (SISSA), Via Bonomea 265, 34136 Trieste, Italy}

\begin{abstract}
We present a numerical investigation of the size scaling of static friction in incommensurate 
two dimensional contacts performed for different lateral loading configurations. 
Results of model simulations show that both the absolute value of the force $F_s$ and the 
scaling exponent $\gamma$ strongly depend on the loading configuration adopted to drive the slider along the
substrate. 
Under edge-loading a sharp increase of static friction is observed above a critical size
corresponding to the appearance of a localized commensurate dislocation.
Noticeably, the existence of sublinear scaling, which is a fingerprint of superlubricity, 
does not conflict with the possibility to observe shear-induced localized commensurate regions 
at the contact interface.
Atomistic simulations of gold islands sliding over graphite corroborate these findings suggesting that similar
elasticity effects should be at play in real frictional contacts.
\end{abstract}

\maketitle

\section{Introduction}\label{sec.intro}

Sliding friction phenomena abound in nature, spanning, in disparate areas, vastly different scales of length, 
time, and energy~\cite{VanossiRMP2013}. However, despite their fundamental and technological importance, several 
key physical aspects of mechanical dissipative dynamics are not yet fully understood.
This is mostly due to the complexity of highly out-of-equilibrium nonlinear processes occurring across a buried 
sliding interface~\cite{Urbakh2004}. Quite recently new avenues of research are being pursued and new discoveries 
are being made especially at the microscopic scales~\cite{Urbakh2010,Manini2017}.
In particular, dry solid/solid contacts displaying extremely low values of static and kinetic friction 
are attracting more and more attention due to their great physical and technological interest, e.g., to 
significantly reduce dissipation and wear in mechanical devices functioning at various scales. Unlike 
standard lubricants, solid lubrication arises from incommensurability of rigid interfaces causing effective 
cancellation of interfacial interactions. This phenomenon, often termed structural superlubricity, has been 
demonstrated in a number of experiments at nano- and micro-scales~\cite{Hirano1997,Dienwiebel2004,Filippov2008,Feng2013,Liu2012,Koren2015,Kawai2016,Liu2017,Dietzel2013,Cihan2016,Pierno2015,Bohlein2012,Bylinskii2015,Bylinskii2016,Kiethe2017}, yet its robustness and its upscale to the macroscopic world remain a challenge. The conditions 
of the persistence of superlubricity and the mechanisms of its failure are therefore cast as key questions to 
be addressed at both fundamental and technological levels.

Even in clean wearless friction experiments with perfect atomic structures the interface elasticity itself 
may hinder superlubricity by introducing new energy dissipation channels~\cite{Muser2004,Ma2015,Benassi2015,Sharp2016}.
Recently, elasticity effects in contact area scale-up have shown to be critical for the breaking
of ultralow frictional states when the contact size exceeds the core width of interfacial 
dislocations in crystalline interfaces~\cite{Sharp2016}. Experimental evidences 
have been provided by measuring the sliding friction force of amorphous antimony particles on 
MoS$_2$~\cite{Schirmeisen2017}. Again in connection to elasticity, the way that a driving local stress 
is applied to a slider, as typically occurs in proximal-probe nano-manipulation measurements, 
is expected to influence its tribological response as well. Indeed, within the idealized framework of a simple 
one-dimensional (1D) edge-driven Frenkel-Kontorova (FK) model~\cite{Ma2015,Benassi2015}, the abrupt occurrence
of a striking boost in the dissipated energy during sliding (kinetic friction) has been observed 
above a critical contact length. 

In the present study, we consider two-dimensional (2D) superlubric 
elastic contacts, where even in the absence of defects a static friction arises due to finite-size edge 
effects~\cite{Varini2015}. In order to unveil the driving-induced mechanisms leading to the substantial 
elimination of superlubricity in a tribologically meaningful 2D geometry, we consider a 
FK-like modeling, highlighting the frictional phenomenology, and a comparative, more realistic, 
adsorbate/substrate atomistic simulation approach, mimicking the frictional interface of gold islands deposited 
on graphite.
We found that the magnitude of the static friction and its scaling with the contact area strongly depend on the 
external lateral loading configuration. Large differences in the static friction observed between edge and uniform  
driving are tightly linked to the pre-slip strain distribution at the frictional interface. In 
particular, load induced formation of localized commensurate regions can add to the edge contribution causing 
a sharp increase of the static friction force. 

\begin{figure}[tb]\label{fig1:schematics}
  \centering
  \includegraphics[width=0.9\columnwidth]{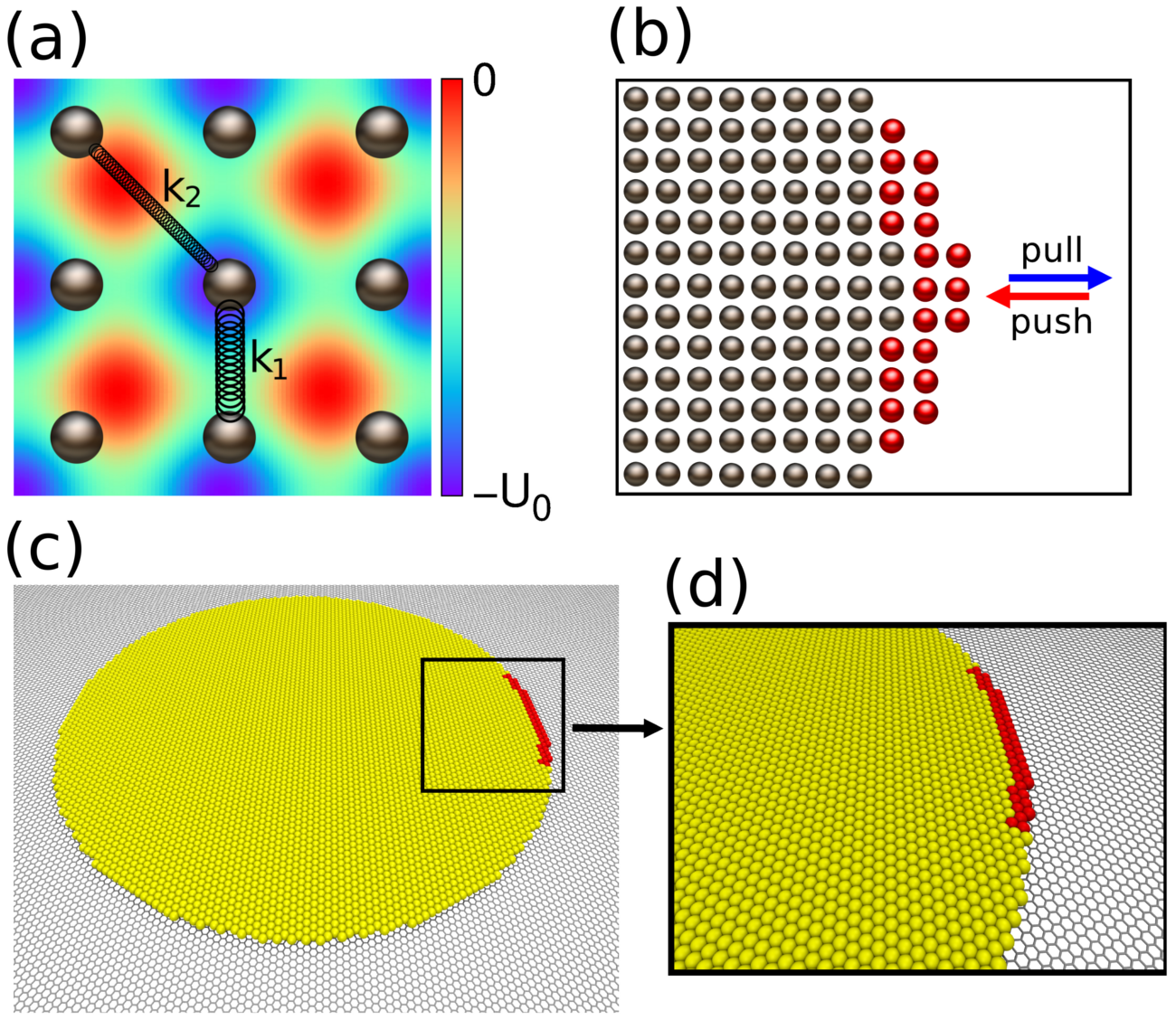}
  \caption{\small (a) The 2D FK model adopted in simulations: a harmonic square lattice of identical point-mass
   particles interacting with a mismatched square substrate-potential of strength $U_0$.
   (b) Schematics of the edge of a circular island in the 2D FK model, showing in red the
   particles to which the external force is applied during the edge-driving pulling/pushing protocols.
   (c) A circular gold island deposited over a graphene monolayer used in the atomistic simulations. (d) Detail 
   of the region where the external force is applied during the edge-driving protocols (red atoms).}
\end{figure}

\section{FK modeling and results}

\subsection{System and Method}\label{sec.method}

In our 2D FK model schematically sketched in Fig.~\ref{fig1:schematics}a, point-like particles of mass $m$ form 
a square lattice of period $a$. Each particle interacts with its four nearest-neighbors via a potential
\begin{equation}
V_{1}(r_{ij})=
\begin{cases}
\frac{k_{1}}{2}(r_{ij}-a)^2 & \text{if $r_{ij} > R_{cut}$;}\\
\alpha+\beta/r_{ij}^{12} & \text{if $r_{ij} \le R_{cut}$;}
\end{cases}
\end{equation}
where $r_{ij}=|{\bf r}_i-{\bf r}_j|$ is the distance between particle $i$ and particle $j$. To get a closer
description of a solid state system we introduced a cut-off distance $R_{cut}$, below which the interaction
turns from harmonic to purely repulsive. The latter term models the Pauli repulsive forces between
electron clouds of real atoms. The constants $\alpha$ and $\beta$ are chosen to ensure the
continuity of $V_1(r)$ up to its first derivative at $r=R_{cut}$.
The isostatic instability under shear forces of the square lattice is eliminated by including a second harmonic term
accounting for the interactions between next nearest-neighbors:
\begin{equation}
V_{2}(r_{ij})=\frac{k_{2}}{2}(r_{ij}-l_{next})^2,
\end{equation}
where $l_{next}=a\sqrt2$ is the next nearest-neighbor equilibrium distance, and the spring constant $k_2$
is chosen to give $k_1/k_2=2$, resulting in a 2D Poisson's ratio of $\sigma_{2D}=1/3$. Values adopted
to perform the simulations are: $a=1$, $m=1$, $R_{cut}=0.85$, $k_1=10$, $k_2=5$.

A rigid crystalline substrate is modeled as a two dimensional periodic potential of square symmetry,
strength $U_0$, and periodicity $\lambda_{sub}$:
\begin{equation}
U({\bf r_i})=-\frac{U_0}{4}\left(2+\cos\frac{2\pi x_i}{\lambda_{sub}}+\cos\frac{2\pi y_i}{\lambda_{sub}}\right).
\end{equation}
We used $\lambda_{sub}=(1+\sqrt{5})/3\approx1.08$, resulting in an overdense interface, where the particle density is 
larger than the density of substrate potential minima. We chose a value of $U_0=0.075$ corresponding to an interfacial 
stiffness $U_0\pi^2/\lambda_{sub}^2\approx0.6$, which is more than one order of magnitude smaller than the internal 
stiffness $k_1=10$ of the monolayer, ensuring that the system is well within the superlubric regime. 
This is usually the case in nano-manipulation experiments~\cite{Dietzel2008,Dietzel2013,Cihan2016}.
This condition has been carefully checked in our model by measuring the static friction force of
infinite monolayers in periodic boundary conditions, as detailed in Section~S1 of Supplemental
 Material (SM)~\cite{SM}. Together with the geometrical lattice mismatch, the ratio of the internal and interfacial 
stiffnesses is directly related to the characteristic core width $b_{core}$ of interfacial edge dislocations. 
This parameter was introduced to define a critical contact size above which 
elasticity becomes important and leads to local locking into the commensurate state~\cite{Sharp2016}. 
For our 2D FK system we estimated $b_{core}\approx68$\, $a$ (see Section~S2 of SM for details).

At rest, the total energy of a system of $N_p$ particles is
\begin{equation}
E(\{{\bf r}_i\})=\sum_{i=1}^{N_p} U({\bf r_i}) + \sum_{\langle i,j\rangle} V_1(r_{ij}) + \sum_{\langle i,j\rangle_{next}} V_2(r_{ij})
\end{equation}
where $\langle i,j\rangle$ and $\langle i,j\rangle_{next}$ indicate summations over all distinct nearest- and 
next nearest-neighbor pairs, respectively. The dynamics of the system is obtained assuming zero temperature
by solving the $N_p$ equations of motion
\begin{equation}
\label{eq:Newton}
m \ddot{\bf r}_i = -\nabla_{i}E(\{{\bf r}_i\})-\eta{\bf v}_i+{\bf F}_{ext},
\end{equation}
where $\eta$ is a damping coefficient accounting for the dissipation of kinetic energy of the particles in 
the island into the substrate and ${\bf F}_{ext}$ is the external driving force.
The equations have been solved numerically with an adaptive Runge-Kutta integrator of the fourth order.
Throughout the paper all results obtained within the 2D FK model are expressed in simulation units,
i.e., as obtained directly from the simulations performed at our chosen parameters,
without any further conversion. Distances are understood to be in units of $a$.

\subsection{Simulation protocols}\label{sec:protocol}

We considered circular islands of increasing radius $R=15$ -- $350$, containing up to $N_p\simeq 3.8\cdot10^5$ 
particles.
The starting configurations were generated adopting the following protocol. For each size we considered a
set of starting angular orientations $\theta$ between the crystalline axis of the island and of the substrate.
We explored values between zero and five degrees, which include the Novaco-McTague
orientation $\theta_{NM}$ ($ \approx 2.5^\circ $, within the chosen parametrization) predicted to be energetically
favorable in the limit of an infinite monolayer~\cite{Novaco1977}. The positions of all particles in the islands 
were fully relaxed via damped dynamics described by the equations of motion \eqref{eq:Newton} in the absence of 
external force. At each size we selected the relaxed configuration corresponding to the minimum value of the total 
energy. During relaxation the islands tended to rotate
from the initial orientation towards the optimal misalignment $\theta_{opt}$ that minimizes the energy.
For the largest sizes considered, where edge effects are minimized, we observed $\theta_{opt}\approx\theta_{NM}$, 
as predicted by linear response theory~\cite{Novaco1977}.

The optimized configurations were used as starting
points for the measurement of the static friction force $F_s$. The latter has been calculated using three different 
lateral loading configurations: edge loading, which mimics the sideway pushing and pulling by a 
tip of atomic force microscope (AFM), and an uniform driving, which can be realized when the slider is attached 
to a sufficiently rigid moving stage or subject to inertial forces as in quartz crystal microbalance experiments. 
To model the edge loading, we selected a small region at the edge of our islands
comprising $\sim70$ particles, to which we apply the external force $F_{ext}$ (see Fig.~\ref{fig1:schematics}b).
An uniform driving is achieved applying $F_{ext}$ to all particles.
The static friction force is evaluated using an adiabatic protocol during which $F_{ext}$
is increased in steps of $\Delta F=0.0001$ -- $0.005$, much smaller 
than the single particle depinning force $F_{s1}=\pi U_0/2\lambda_{sub}\simeq 0.11$. For each value 
of $F_{ext}$, the positions of all particles are relaxed via damped dynamics, using a damping $\eta=0.5$. 
Depinning was detected by monitoring the displacement of the island's center-of-mass. We note that,
as long as inertial effects are negligible during relaxation, the value of the static friction force is 
independent of the choice of the damping coefficient. To check the reliability of our protocol, we repeated some 
relaxations with the more sophisticated FIRE optimization algorithm~\cite{Bitzek2006}, which
yielded the same results. The data presented here were obtained applying the external force in the 
direction defined by $\theta_{opt}$, corresponding to a high symmetry lattice
direction of the slider. We checked that the main results are qualitatively independent of the direction of 
application of the external force.

The applied external force induces strain deformations within the slider. These distortions occur mainly
in the direction of the external force $F_{ext}$, while the Poisson effect accounts for relatively
smaller deformations in the perpendicular direction. We focus on the first, and we quantify
them by computing the local average distance $\delta_{loc}({\bf r}_i)$ of each particle
$i$ from its nearest-neighbors lying along the direction of the applied force.
Two dimensional color maps of $\delta_{loc}$ are produced to gain insights on the
driving-induced strain distribution. One dimensional plots are extracted by averaging
$\delta_{loc}$ in the direction perpendicular to $F_{ext}$,
$\delta(x)=\langle\delta_{loc}({\bf r}_i)\rangle_{\perp}$.
Both $\delta_{loc}$ and $\delta$ are expressed in units of the substrate periodicity $\lambda_{sub}$,
so to easily detect local commensuration when $\delta_{loc}/\lambda_{sub} = 1$.
\begin{figure}[tb]\label{fig.2:scaling_FK}
  \centering
  \includegraphics[width=0.9\columnwidth]{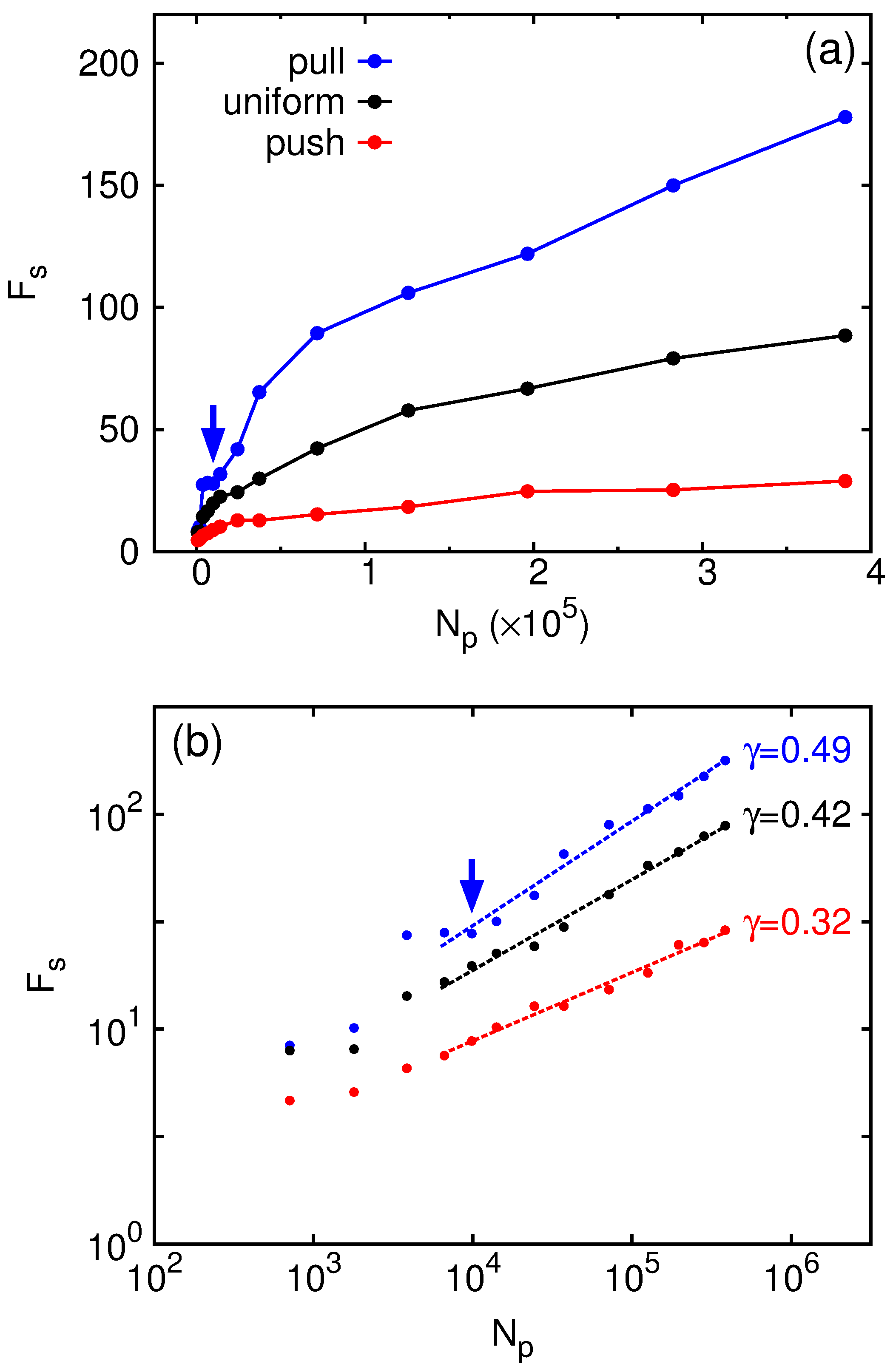}
  \caption{\small (a) Size scaling of the static friction force. 
                  (b) A linear fit of the data in a Log-Log scale (dashed lines) 
                  yielded the scaling exponents $\gamma$ reported beside each curve. Blue, black,  
                  and red lines are results obtained under pulling, pushing, and uniform driving, respectively. 
                  Arrows indicate the critical size $N_p^*$ for the nucleation of a local commensurate dislocation
                  in the case of pulling.
          }
\end{figure}
\begin{figure}[tb]\label{fig.3:maps_FK}
  \centering
  \includegraphics[width=\columnwidth]{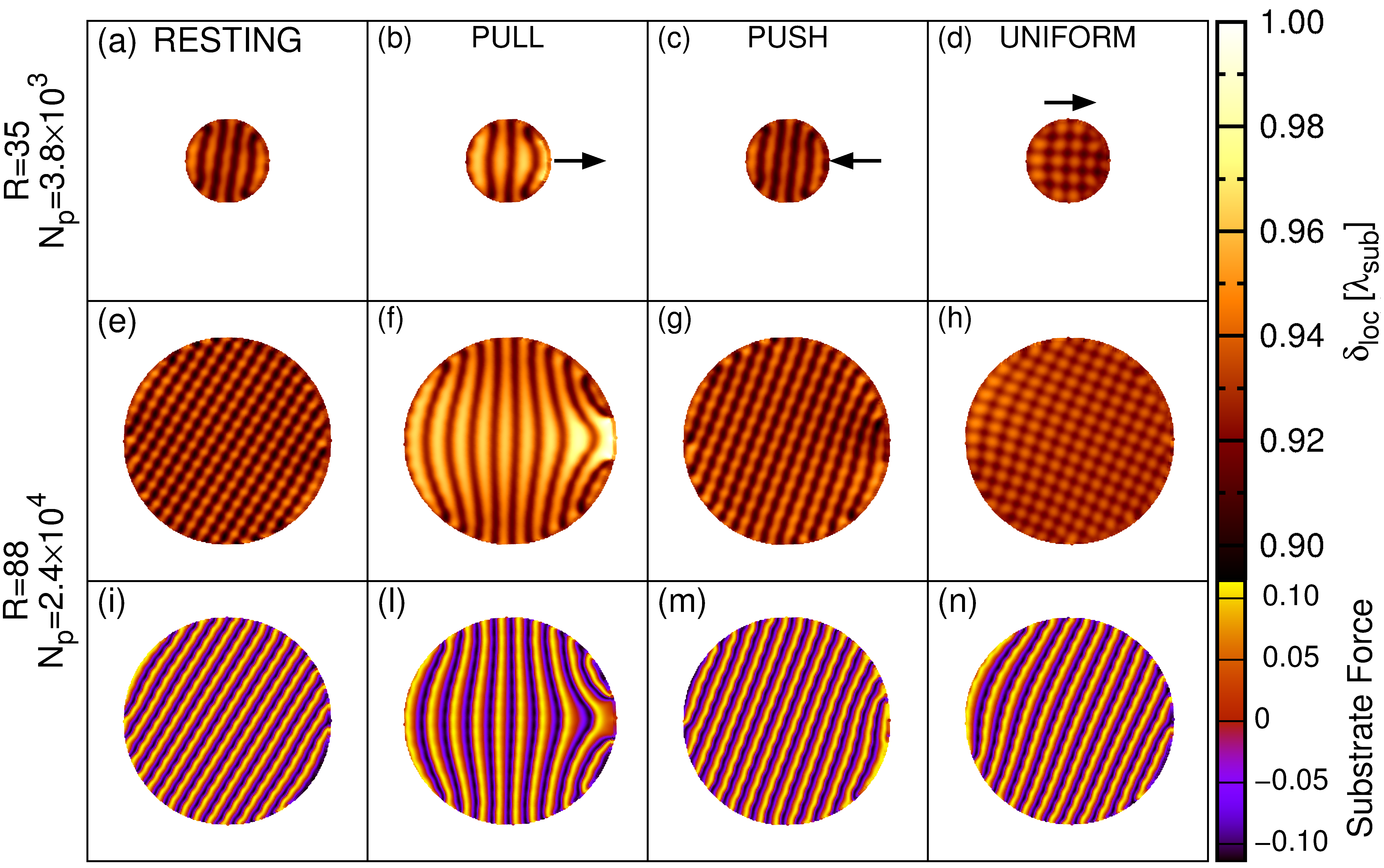}
  \caption{\small (a)-(h) Colored maps showing the local mismatch, $\delta_{loc}$, of the circular islands in the 
           2D FK model simulations. Values of $\delta_{loc}=1$ indicate local commensuration to 
           the substrate. Panels (a)-(d) show results for a size of $N_p\simeq3.8\cdot10^{3}$ particles, 
           while panels (e)-(h) correspond to $N_p\simeq2.4\cdot10^{4}$.
           (i)-(n) Color maps of the substrate potential force (projected along the direction opposite 
           to that of the external force) in the island of size $N_p\simeq2.4\cdot10^{4}$. Positive values 
           correspond to regions resisting the motion. Within the bulk the substrate forces alternate in 
           sign, summing up to nearly zero, and pinning originates mainly at the edges.
           From left to right, snapshots are shown obtained at rest ($F_{ext}=0$, panels (a),(e),(i)), 
           and at an applied external force just below the static friction force value, $F_{ext}\lesssim F_s$, 
           under edge-pulling (panels (b),(f),(l)), edge-pushing (panels (c),(g),(m)), and uniform driving
           (panels (d),(h),(n)). Arrows indicate the direction of $F_{ext}$ in the three protocols.
          }
\end{figure}
\begin{figure}[tb]\label{fig.4:localcomm_FK}
  \centering
  \includegraphics[width=0.9\columnwidth]{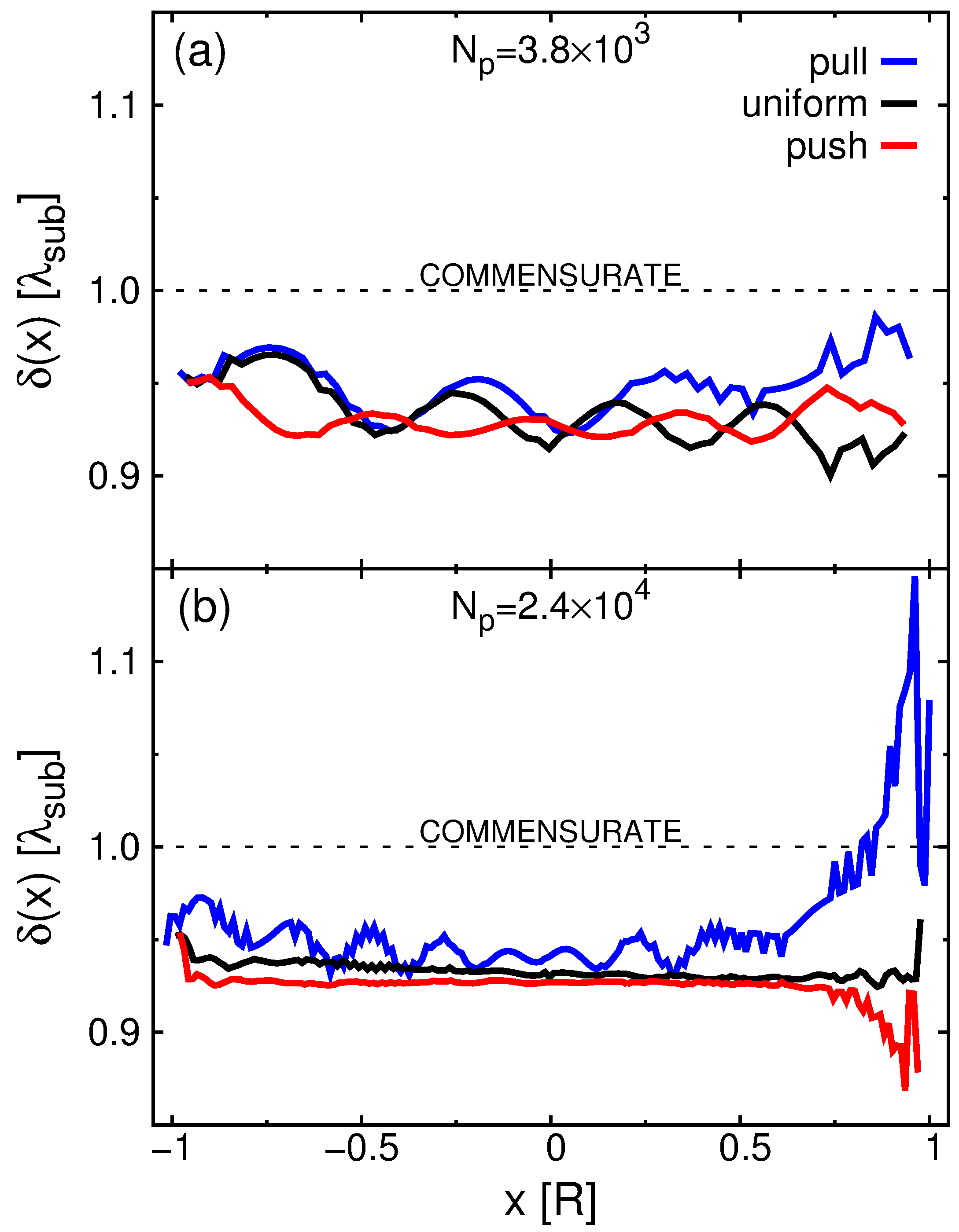}
  \caption{\small The averaged local mismatch, $\delta(x)$, inside the circular islands
           of the 2D FK model simulations. Values of $\delta(x)=1$ indicate local commensuration to the
           substrate. Panels (a) and (b) show results for two islands' sizes of $N_p\simeq3.8\cdot10^{3}$
           and $N_p=2.4\cdot10^{4}$ particles, corresponding to radii of $R=35$ and $R=88$, respectively. 
           Blue, red, and black lines are obtained from the analysis of the configurations relaxed at 
           an external force just below the static friction force, $F_{ext}\lesssim F_s$, under edge-pushing, 
           edge-pulling, and uniform driving, respectively.
          }
\end{figure}
\subsection{Results: Size scaling and driving-induced elasticity effects}
Figure~\ref{fig.2:scaling_FK}, the main result of this study, reports the 
size dependencies of the static friction force calculated for three different lateral loading configurations:
edge-pulling, uniform driving, and edge-pushing. We extract three main observations:
(i) Static friction obeys a sublinear size scaling law, $F_s\propto N_p^\gamma$ with $\gamma<1$, 
independent of the driving protocol (see Fig.~\ref{fig.2:scaling_FK}b).
(ii) At all sizes we consistently observe $F_s^{pull}>F_s^{uniform}>F_s^{push}$. The effect 
of non-uniform driving is indeed quite large: the static friction force measured for
edge-pulling being more than twice that obtained under uniform shearing, and almost one order of magnitude larger 
than the value measured for edge-pushing.
(iii) More surprisingly, the scaling exponent itself is found to depend significantly on the adopted lateral loading, 
and shows the same trend, $\gamma_{FK}^{pull}>\gamma_{FK}^{uniform}>\gamma_{FK}^{push}$, of the corresponding  
static friction force values.

To understand the physical origin of this behavior we analyzed the pre-slip strain distribution at the 
interface. Figure~\ref{fig.3:maps_FK}a-h reports the colored maps of $\delta_{loc}$ for two
sizes, comparing the configurations at rest ($F_{ext}=0$, see
Fig.~\ref{fig.3:maps_FK}a,e) with the ones relaxed in presence of an external force just below the static friction 
threshold, $F_{ext}\lesssim F_s$ (see Fig.~\ref{fig.3:maps_FK}b-d,f-h). At all sizes
the strain distribution is strongly dependent on the lateral loading. For the overdense case considered, 
pulling induces elongations of the inter-particle distances in the direction of the applied external force, 
while pushing tends to reduce the bond-lengths. In both cases the largest distortions are localized near 
the edge region to which the external force is applied (see Fig.~\ref{fig.3:maps_FK}b,c,f,g). An uniform 
shearing generally yielded mild compressions at the leading edge and somewhat larger elongations at the trailing edge
(see Fig.~\ref{fig.3:maps_FK}d,h, and the full nearest-neighbor distances distributions reported in Fig.~S2 of SM).
Elasticity effects are particularly large for the edge-pulling protocol. There,
as the size of the slider increases, the local bond-length approaches values that grow closer and closer to 
the substrate periodicity, $\delta_{loc}/\lambda_{sub}\rightarrow 1$ (see Fig.~\ref{fig.3:maps_FK}b,f).

In finite 2D superlubric contacts, locking to the substrate originates from lower-coordinated atoms 
that soften the slider at its border, hence the static friction scaling is expected to increase
at most as $\gamma_{max}=1/2$. In our system, evidences of this behavior can be observed from the colored 
maps of the substrate forces acting on the slider (see Fig.~\ref{fig.3:maps_FK}i-n). In agreement with 
previous work~\cite{Varini2015} we actually found values of $\gamma<1/2$ (see Fig.~\ref{fig.2:scaling_FK}b), 
indicating that pinning originates from a subset of points that grows sublinearly even with respect to the edge. 
Defining $F^*$ as the force needed to nucleate a localized commensurate dislocation, there must 
exist a corresponding threshold size $N_p^*$ above which $F^*\leq F_s(N_p)$ and local matching, namely 
$\delta_{loc}({\bf r})/\lambda_{sub}=1$ for some {\bf r}, should occur before the onset of motion. 
We found that this is indeed the case, as demonstrated in
Fig.~\ref{fig.4:localcomm_FK}. There we show the average nearest-neighbor distance $\delta(x)$ measured along
the direction of the external force, for two distinct island sizes. For $N_p=3.8\cdot10^3<N_p^*\approx 10^4$, 
the island is incommensurate everywhere, independent of the driving protocol. At $N_p=2.4\cdot10^4>N_p^*$ a 
localized commensurate dislocation is formed close to the pulling region. The same behavior was observed at 
each value of $N_p>N_p^*$, up to the largest size investigated. On the other hand, for edge-pushing and
uniform lateral loading, we found no evidence of nucleation of a commensurate dislocation, up to the largest
value of $R=350$ ($\sim5$\,$b_{core}$) considered. The emergence of commensurability is expected to affect 
the measured value of the static friction force, and indeed, for edge-pulling we found a sharp increase just above 
$N_p\approx N_p^*$ (see arrow in Fig.~\ref{fig.2:scaling_FK}a). 

Coming to the scaling exponents, in Fig.~\ref{fig.2:scaling_FK}b we report the results 
of a fit of the simulations' data. In our overdense contact, pushing promotes incommensurability by 
compressing the slider lattice spacing $a$. Correspondingly, we observed the smallest value of 
$\gamma_{FK}^{push}=0.32\pm0.01$. The uniform shearing yielded an intermediate value $\gamma_{FK}^{uniform}=0.42\pm0.01$, 
in agreement with the observation of more enhanced bond elongations, compared to compressions, taking place 
at the interface (see Fig.~S2 of SM). We note that a similar value of $\gamma^{uniform}\approx0.4$ was
previously reported in atomistic simulations of circular Kr islands sliding on Pb(111)~\cite{Varini2015}.
Finally, the largest value $\gamma_{FK}^{pull}=0.49\pm0.02$ is obtained for edge-pulling, reflecting the pinning 
effect of the localized commensurate dislocation, that adds to the edge contribution (see also the force map 
of Fig.~\ref{fig.3:maps_FK}l). 

It is worth stressing that despite the presence of a commensurate dislocation, $F_s$ obeys a sub-linear 
scaling law even under edge-pulling, indicating that the commensurate area does not grow linearly with 
the contact size.
Thus, our simulations demonstrate that the emergence of a localized commensurate region is not
incompatible with the observation of a static friction sublinear scaling, which is usually considered a 
fingerprint of superlubricity.
This conclusion differs from that of the work by Sharp et al.~\cite{Sharp2016}, where, within 
a similar simplified elastic model under the uniform loading, a transition from a sublinear to linear 
scaling of $F_s$ has been predicted for the contact size corresponding to the appearance of commensurate 
dislocations. In that case, the linear scaling of $F_s$ is determined by the increase of the number of dislocations 
that can be stabilized at the (quasi-statically) sliding interface~\cite{Hurtado1999}.
It should be noted that, at least for some set of parameters, the results of simulations 
in Ref.~\cite{Sharp2016} show the presence of localized commensurate dislocations already in the starting 
configurations at rest, suggesting that the contacts were not in the superlubric regime 
(see discussion in Section~S1 of SM). In our case, an externally applied non-uniform driving force is necessary 
for the appearance of a commensurate dislocation. Moreover, during sliding the 
dislocation remains localized near the edge, and no others are nucleated across the contact (see Fig.~S3 of SM).
\section{Atomistic MD simulations}\label{sec:gold}
\subsection{Model: Au(111) over graphite}\label{subsec:model-Au}
The comparative atomistic MD investigation deals with gold islands deposited on a graphitic substrate.
We have considered Au(111) islands of circular shape (see Fig.~\ref{fig1:schematics}c,d),
fully described atomistically by EAM potential~\cite{EAM}, with a spacing parameter $a = 4.63$\,\AA, and deposited
on a graphene monolayer kept rigid in its bulk configuration, corresponding to a lattice constant 
$\lambda_{sub} = 4.26$\,\AA. Au-C interaction has been mimicked by a 12-6 Lennard-Jones potential, 
with $\epsilon = 44$\,meV, $\sigma = 2.74$\,\AA, and a cutoff radius of 7\,\AA.
In order to reduce the critical island size at which its elasticity affects friction, we have used the 
same $\sigma$ but $\epsilon$ twice that of Lewis et al.~\cite{Lewis2000}.
We have considered Au island sizes with radii in the range $R=6$ -- $40$\,nm, the largest corresponding 
to about $8\cdot10^4$ Au atoms, adsorbed on a graphene sheet of 88$\times$88\,nm$^2$ ($\sim3\cdot10^5$ C atoms) 
with applied periodic boundary conditions. The characteristic core width of interfacial edge dislocations for 
this system is of the order of $b_{core}\approx18$\,nm$=39$\,$a$ (see Section~S2 of SM).
After damped-dynamics relaxation we obtain an optimum alignment angle $\theta_{opt}\simeq1.7^\circ$, as 
similarly observed in earlier work~\cite{Guerra16}. Dynamics was simulated using a velocity-Verlet integrator 
with a time step of 1\,fs, and a viscous damping of $\eta=0.1$\,ps$^{-1}$. The protocols adopted for the initial
relaxation and the subsequent measurement of the static friction force are the same as outlined in 
Section~\ref{sec:protocol}.
\begin{figure}[tb]\label{fig.5:maps_gold}
  \centering
  \includegraphics[width=\columnwidth]{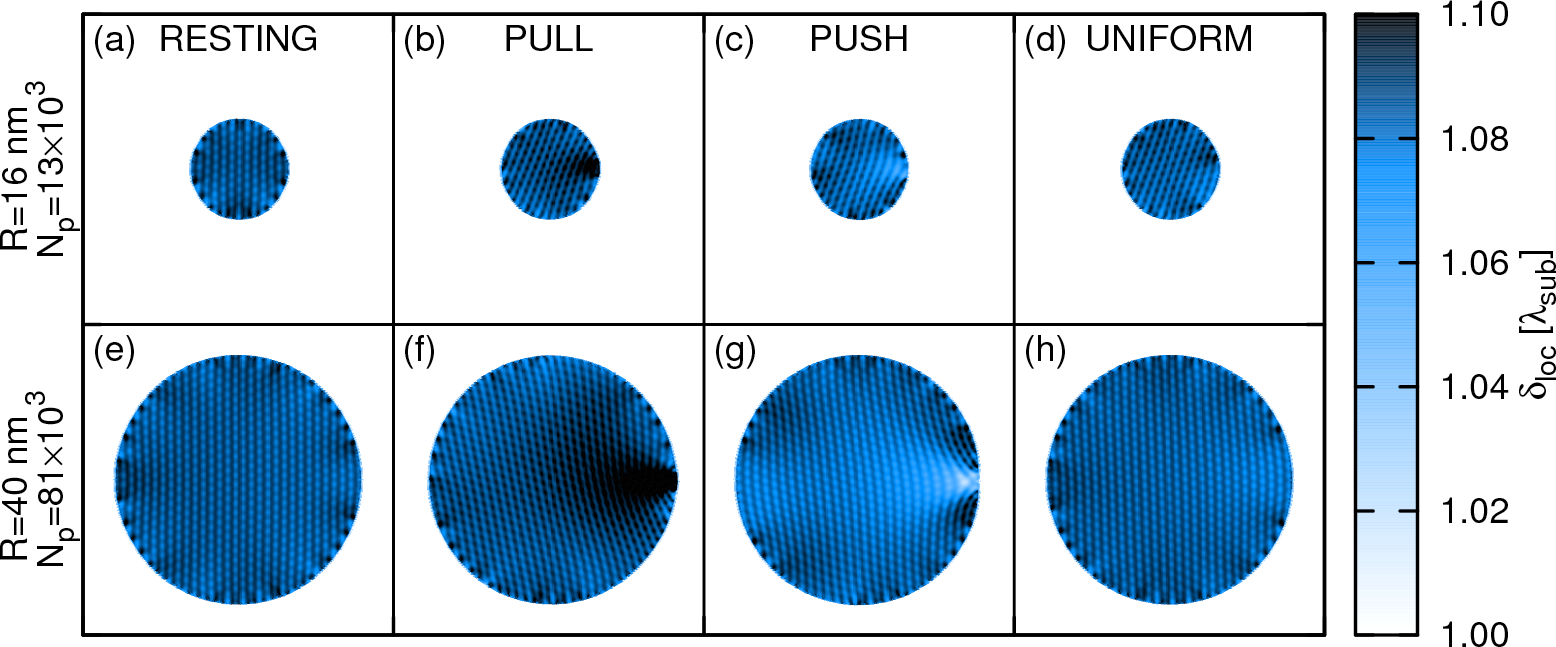}
  \caption{\small Local mismatch maps for the gold-graphene system at two different island size, $R=16$\,nm (a-d)
           and $R=40$\,nm (e-h). Beside the resting (no external force) case, configurations for pulling, pushing,
           and uniform driving are reported with an applied force just below the depinning threshold. Local
           commensuration is achieved in the white region.}
\end{figure}
\begin{figure}[bt]\label{fig.6:scaling_gold}
  \centering
  \includegraphics[width=\columnwidth]{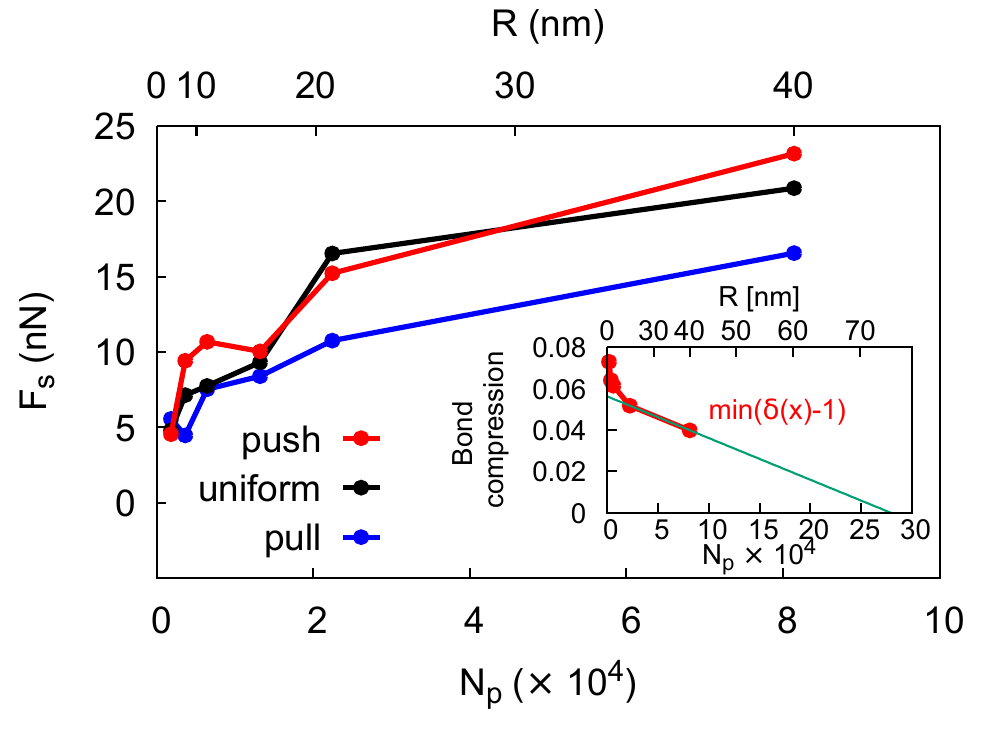}
  \caption{\small Size scaling of the static friction force $F_s$ in the gold-graphene system.
                  Blue, black, and red lines are results obtained during the pulling,
                  uniform, and pushing protocols, respectively. The inset shows the scaling of the
                  maximum bond compression during pushing. A linear extrapolation is used to obtain an
                  approximate estimation of the critical size for the appearance of a local commensurate dislocation.
          }
\end{figure}
\subsection{Results: interface strain distribution and size scaling}\label{subsec:resultsAu}
Contrary to the FK system, the existent mismatched ratio between
gold and graphene defines an underdense contact geometry with $\lambda_{sub} < a$.
Thus, adsorbate/substrate commensuration is expected to be favored by a sideway pushing instead of pulling.
Figure~\ref{fig.5:maps_gold} reports the colored strain maps for two island sizes and
the different driving protocols previously considered, where close commensurate regions
with the underneath substrate are highlighted by brighter tones.
While the smaller adsorbate (top panels) does not exhibit significant strain deformations 
regardless of the type of applied shearing, the larger slider (bottom panels) starts to develop,
under sideway pushing, the nucleation of a localized nearly commensurate zone.

Figure~\ref{fig.6:scaling_gold} reports the corresponding size scaling of the static friction force.
We observe $F_s^{pull}<F_s^{push}\approx F_s^{uniform}$, indicating again that pinning to the 
substrate can be reduced by adopting lateral loading configurations that promote incommensurability.
Under edge-pushing and uniform driving, up to the largest size investigated, local commensuration is approached but
never reached. 
Correspondingly, we measured comparable values of the static friction force, as similarly observed in the FK model 
well below $N_p^*$ (see Fig.~\ref{fig.2:scaling_FK}b). A tentative fit of the simulations' data (see Section~S5 of SM)
yielded $\gamma_{gold}^{pull}=0.33\pm0.06$, $\gamma_{gold}^{uniform}=0.39\pm0.05$, and 
$\gamma_{gold}^{push}=0.38\pm0.06$. In analogy with the FK model, we note that 
the scaling exponents follow qualitatively the same trend displayed by the corresponding values 
of the static friction force. Moreover, we found agreement between the values 
of $\gamma_{gold}^{pull}\approx\gamma_{FK}^{push}$ and of $\gamma_{gold}^{uniform}\approx\gamma_{FK}^{uniform}$. 
On the other hand, since local commensuration is never reached in the gold-graphene system under sideway pushing, we 
observe $\gamma_{gold}^{push}<\gamma_{FK}^{pull}$.
A rough estimate of the critical size for the appearance of a commensurate dislocation in the gold islands is given in the inset of Fig.~\ref{fig.6:scaling_gold}. 
There we show the size-dependence of 
the minimum distance from achieving local commensuration, ${\rm min}[\delta(x)-1]$, measured for the 
pushing protocol. A linear extrapolation indicates that commensurate dislocations should nucleate above a 
critical island size of $R\gtrsim75$\,nm. The numerical inspection of such computationally demanding limit
goes beyond the scope of the study.
Nevertheless, the close similarity of the strain maps of Fig.~\ref{fig.5:maps_gold}
with those of Fig.~\ref{fig.3:maps_FK}, and the differences between the values of the static friction force 
highlighted in Fig.~\ref{fig.6:scaling_gold} suggest that a similar 
boost in the static friction is to be expected in sufficiently large real contacts as well.

\section{Discussion and Conclusions}\label{sec.conclusions}

We carried out a detailed numerical investigation of the static friction force $F_s$ arising in 
crystalline nano-contacts for three different lateral loading configurations. We focused on incommensurate 
superlubric interfaces, for which a sublinear scaling of $F_s\propto A^{{\gamma}<1}$ with respect to the 
contact area $A$ has been predicted theoretically~\cite{deWijn2012,Varini2015}, and observed 
experimentally~\cite{Dietzel2013,Cihan2016}. 
Results obtained within the framework of a two 
dimensional Frenkel Kontorova model~\cite{Braun1998} demonstrate that both the absolute value $F_s$ and the 
scaling exponent $\gamma$ depend on the lateral loading configuration adopted to drive the slider along the
substrate. Specifically, we observed variations of nearly 
one order of magnitude in the measured value of $F_s$ upon changing the direction of the applied shearing 
force during sideway pushing/pulling of circular islands. On top of that, the scaling exponent $\gamma$ itself was 
found to vary by $\sim 50$\ \%. 

This behavior is tightly linked to the pre-slip strain distribution induced 
at the frictional interface. Larger (smaller) values of ($F_s$,$\gamma$) are observed in concomitance of lateral 
loading configurations that stretch (compress) the slider lattice spacing $a$ towards (away from) 
matching with the periodicity $\lambda_{sub}$ of the underlying crystalline substrate. 
A large boost of static friction is observed when the contact area exceeds a critical value above which the static 
friction force becomes larger than the force $F^*$ needed to nucleate a localized commensurate dislocation near 
the driven edge. Remarkably, independent of the presence or absence of commensurate dislocations, the scaling law 
remains sub-linear, indicating that the commensurate area does not increase proportionally to $A$. 
This result further demonstrates that a sublinear scaling, usually considered a fingerprint of superlubricity,
does not conflict with the possibility to observe shear-induced localized commensurate regions at the contact 
interface. Comparative atomistic simulations of gold nano-sliders deposited 
over a graphitic substrate corroborated these findings thus suggesting that similar effects should be at play in 
sufficiently large real contacts that can be probed in nano-manipulation experiments~\cite{Cihan2016}. 

We mention here that previous works~\cite{Ma2015,Benassi2015} dealing with an edge-driven one dimensional FK model
reported the occurrence of a boost of energy dissipation above a critical length corresponding  
to the nucleation of a localized commensurate region. In that case, the threshold value was determined
by the linear increase with contact size of the viscous-like kinetic friction force needed for attaining 
steady-sliding at a given constant velocity. In superlubric one dimensional contacts of increasing size
static friction instead oscillates around a constant value~\cite{Gigli2017}, hence the creation of a commensurate 
dislocation before the onset of motion is hindered if $F_s<F^*$, which is generally the 
case. Again in connection with elasticity, in a recent work by Sharp et al.~\cite{Sharp2016}, a transition from 
a sublinear to linear scaling of $F_s$ has been predicted for the contact size corresponding to the 
appearance of interfacial commensurate dislocations characterized by a core width $b_{core}$. 
However, in that case the transition was observed only for the case of soft sliders, corresponding to small 
values of $b_{core}/a\lesssim2$. Moreover, local commensuration was reported even in the absence of external 
shearing forces, suggesting that the contacts were not genuinely superlubric. 
On the contrary, the results of our simulations falls in the regime of stiff sliders and a 
large $b_{core}/a\gg1$, 
where local commensuration is observed only under the action of a non-uniform lateral loading. The present 
investigation therefore highlights a new mechanism for nucleation of edge-driving induced commensurate 
dislocations which lead to a significant increase of the static friction force.

In our simulations, the external force $F_{ext}$ was applied to $N_{driven}\sim70$ particles at 
the edge of the slider. More generally, the frictional response of the system will depend on 
the fraction $f_d=N_{driven}/N_p$ of driven particles. For a fixed size $N_p$ of the slider, under sideway
pushing/pulling, the static friction force will tend to the uniform driving value, $F_s\to F_s^{uniform}$ for 
$f_d\to 1$, even though not necessarily in a continuous fashion. We can expect a similar scenario also for the
scaling exponent. Increasing the fraction $f_d$ of driven particles will lead to a crossover of the scaling
exponent $\gamma(f_d)$, from the non-uniform value at small fractions, to the uniform value at larger fractions,
where the stress distribution becomes homogeneous. The way the uniform value is approached will depend on the 
details of the definition of the driven region.

Summarizing, while it has been shown that the static friction force scaling exponent $\gamma$ of incommensurate
crystalline interfaces depends on the geometrical properties of the contact, including the shape of the slider and the
mismatch and symmetry of the contacting lattices~\cite{deWijn2012,Varini2015}, our results prove that $\gamma$
can vary significantly due to elasticity effects. These, in turn, can be characterized in terms of the interfacial 
strain distribution induced by the externally applied shear stress. In particular, in sufficiently large contact
geometry, the possibility that the edge-driving induced nucleation of localized commensurate dislocations leads to 
the breaking of superlubric sub-linear scalings cannot be excluded, and certainly deserves further investigation.
Interestingly, variations of the static friction with the loading
configuration has been found previously for elastic macroscopic contacts~\cite{Ben-David2011,Capozza2012}, where
this effect results from changes in the rupture dynamics at the rough frictional interface.

\section*{Acknowledgments}
D.M. acknowledges the fellowship from the Sackler Center for Computational Molecular 
and Materials Science at Tel Aviv University, and from Tel Aviv University Center for Nanoscience and Nanotechnology.
W.O. acknowledges the financial support from a fellowship program for outstanding postdoctoral researchers from 
China and India in Israeli Universities.
M.U. acknowledges the financial support of the Israel Science Foundation Grant 1316/13.
A.V. and R.G. acknowledge financial support by ERC Grant 320796 MODPHYSFRIC. 
COST Action MP1303 is also gratefully acknowledged.

\end{document}


\title{Supplemental Material for \\ ``Static friction boost in edge-driven incommensurate contacts''}

\author{
  \firstname{Davide} \surname{Mandelli}}
  \affiliation{Department of Physical Chemistry, School of Chemistry,
  The Raymond and Beverly Sackler Faculty of Exact Sciences and The
  Sackler Center for Computational Molecular and Materials Science,
  Tel Aviv University, Tel Aviv 6997801, Israel}
\author{\firstname{Roberto} \surname{Guerra}}
  \affiliation{Center for Complexity and Biosystems, Department of Physics, University of Milan, 20133 Milan, Italy}
\author{\firstname{Wengen} \surname{Ouyang}}
  \affiliation{Department of Physical Chemistry, School of Chemistry,
  The Raymond and Beverly Sackler Faculty of Exact Sciences and The
  Sackler Center for Computational Molecular and Materials Science,
  Tel Aviv University, Tel Aviv 6997801, Israel}
\author{\firstname{Michael} \surname{Urbakh}}
  \affiliation{Department of Physical Chemistry, School of Chemistry,
  The Raymond and Beverly Sackler Faculty of Exact Sciences and The
  Sackler Center for Computational Molecular and Materials Science,
  Tel Aviv University, Tel Aviv 6997801, Israel}
\author{\firstname{Andrea} \surname{Vanossi}}
  \affiliation{CNR-IOM Democritos National Simulation Center, Via Bonomea 265, 34136 Trieste, Italy}
  \affiliation{International School for Advanced Studies (SISSA), Via Bonomea 265, 34136 Trieste, Italy}

\begin{abstract}
\end{abstract}

\maketitle
\section{Aubry transition in the infinite 2D FK model}\label{secS1:Aubry-FK}
The main scope of the present study is to test the conditions under which elasticity effects may
hamper the superlubricity phenomenon in incommensurate contacts under non-uniform drivings.
To model this scenario care must be taken in the choice of the model parameters. In fact, it has been
reported~\cite{Mandelli2015,Mandelli2017,Brazda2018} that infinite two dimensional incommensurate contacts 
undergo an Aubry-like phase transition~\cite{Aubry1983} -- where static friction turns from zero to finite, 
and superlubricity is lost -- when the substrate potential strength $U_0$ (see equation (3) in the main text) 
exceeds a critical value $U_c$. In Ref.~\cite{Mandelli2015} it was shown that a signature of the
Aubry transition in underdense interfaces ($\lambda_{sub}/a<1$) is the appearance of a network of dislocations
separating domains where the monolayer snaps to commensuration with the substrate. Moreover, the value of $U_c$
was found to depend on the misalignment angle $\theta$ between the contacting lattices.

In order to select a value of $U_0<U_c$, safely within the superlubric regime of our interest,
we performed test simulations of infinite interfaces adopting periodic boundary conditions (pbc).
We considered three supercells realizing misalignment angles of $\theta=0^o$, $5^o$, and $\theta_{NM}=2.6^o$.
The latter corresponds to the optimal (Novaco-McTague) angle predicted to be energetically
favorable~\cite{Novaco1977} for our chosen overdense interface lattice mismatch ratio
$\lambda_{sub}/a=(1+\sqrt{5})/3$ ($>1$), and spring constants ratio $k_1/k_2=2$.
The method adopted to construct the supercells is found in Ref.~\cite{Mandelli2015}.
For each supercell we considered increasing values in the
range $U_0=0$ -- $0.1$, and we relaxed the positions of all atoms via damped dynamics. Finally, for each relaxed
configurations, we established the presence/absence of a finite static friction force by applying
a uniform driving, following the procedure outlined in Section II.B of the main text. We used a step
$\Delta F=2\cdot10^{-5}$ to increase adiabatically the external force per atom. We note here that due
to computational limitations we could not investigate smaller values of $F_{ext}$ within reasonable simulation times.

\begin{figure}[!h]\label{figS1:Aubry-PBC}
  \centering
  \includegraphics[width=0.9\columnwidth]{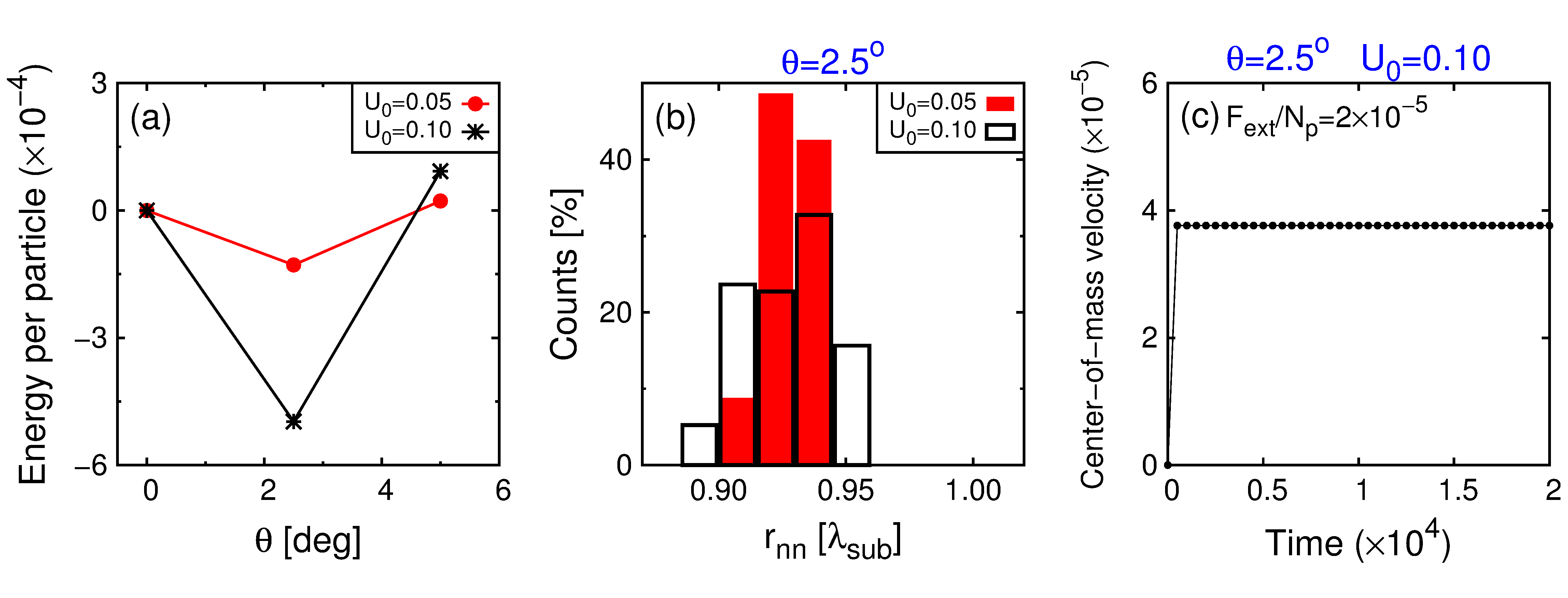}
  \caption{\small (a) Total energy per particle as a function of the misalignment angle $\theta$ in the simulations
           in pbc. The energy of the aligned configuration at $\theta=0^\circ$ is used as reference value.
           Red and black curves are results obtained for two different substrate potential strengths of $U_0=0.05$
           and $U_0=0.1$, respectively. The minimum of the total energy energy coincides with the optimal
           Novaco-McTague angle $\theta_{NM}\approx2.5^\circ$ predicted by linear response theory~\cite{Novaco1977}.
           (b) Distributions of the nearest-neighbor distances $r_{nn}$ (in units of the substrate periodicity
           $\lambda_{sub}$) obtained in the simulations in pbc at $\theta=\theta_{NM}\approx2.5^\circ$.
           Filled red boxes and empty black boxes are results for two different substrate potential strengths of
           $U_0=0.05$ and $U_0=0.1$, respectively. Both distributions are narrow and peaked around the
           equilibrium value, $a/\lambda_{sub}\approx0.93$. Local commensuration ($r_{nn}=1$) is absent.
           (c) The velocity of the center-of-mass of the monolayer in pbc at misalignment
           $\theta=\theta_{NM}\approx2.5^\circ$, and substrate potential strength $U_0=0.1$, sliding subject
           to a small external force of $F_{ext}/N_p=2\cdot10^{-5}$.
           After an initial transient, steady-state is reached and the monolayer slides at constant
           velocity, proving that static friction is vanishingly small. For comparison, note that the
           single particle static friction force $F_{s1}=\pi U_0/2\lambda_{sub}\simeq 0.15$ is four
           orders of magnitude larger than the smallest external force (per particle) considered here. All simulations
           were performed adopting the following set of parameters of our 2D FK model: $a=1$, $k_1=10$, $k_2=5$,
           $\lambda_{sub}\approx(1+\sqrt{5})/3$. For the sliding simulations we adopted
           a viscous damping $\eta=0.5$, which is necessary in order to reach steady state.
          }
\end{figure}

Up to $U_0=0.1$, results show that:
(i) The orientation $\theta = \theta_{NM}$ is
energetically favorable (see Fig.~\ref{figS1:Aubry-PBC}a below), demonstrating that the system is in the linear
response regime expected for weak substrate corrugation strengths~\cite{Novaco1977};
(ii) The relaxed configurations at rest ($F_{ext}=0$) do
not show any localized commensurate dislocation, the final distribution of the nearest neighbor distances
being narrow and peaked around the equilibrium value $a/\lambda_{sub}\approx0.93$
(see Fig.~\ref{figS1:Aubry-PBC}b);
(iii) At each misalignment angle, all the monolayers are free to slide already at the smallest value of the applied
external force, $F_{ext}/N_p=2\cdot10^{-5}$ (see Fig.~\ref{figS1:Aubry-PBC}c).

We conclude that the value of $U_0=0.075$ adopted in the simulations discussed in the main text is reasonably
below the critical threshold $U_c$ for the breakdown of superlubricity induced by the
Aubry transition. This ensures that the bulk of the circular islands effectively model a superlubric contact,
and they would move freely under the action of any arbitrarily small external force if it was not for the pinning
effects introduced by the presence of the edges.
\newpage
{\white Hello}
\vfill
\vfill
\clearpage
\vfill
\section{Estimation of $b_{core}$ in the 2D FK model and in the atomistic simulations}\label{secS2:b-core}
The core width $b_{core}$ of interfacial edge dislocations in our 2D FK model is given by \cite{Sharp2016}
\begin{eqnarray}\label{eq.bcore.Robbins}
b_{core} & = & a G/\tau_{max}\\
\tau_{max} &=& \pi U_0/2\lambda_{sub}^3 \nonumber
\end{eqnarray}
where $a$ and $G$ are the lattice spacing and shear modulus of the slider, and $\tau_{max}$ is the maximum shear
stress imposed by the substrate potential. For the values of $U_0=0.075$ and $\lambda_{sub}=(1+\sqrt5)/3$ used
in the simulations one has $\tau_{max}\simeq0.094$. We computed the shear modulus using the
relation $G=E/2(1+\nu)$~\cite{Jasiuk1994}, where $E$ and $\nu$ are the 2D Young modulus and Poisson's
ratio, respectively. For our chosen square lattice of spring constants $k_1=10$, $k_2=5$, and lattice spacing
$a=1$, one has $E=k_1(1+\sqrt2/2)$, $\nu=1/3$, $G\simeq6.402$, yielding $b_{core}\approx 68$ $a$.

For our considered case of gold islands sliding over graphene Eq.\ \ref{eq.bcore.Robbins} becomes
\begin{eqnarray}
     b_{core} & = & a_x G/\tau_{max} \\
 \tau_{max} & = & F_x^{max}/(b_x \cdot b_y) \nonumber
\end{eqnarray}
where $a_x=4.63$\,\AA\ is the gold lattice spacing along loading direction, $F_x^{max}=62$\,meV/\AA\ is the  maximum force along $x$ imposed by graphene on an Au atom placed at $z=2.7345$\,\AA\ (the average $z$ value of a large resting island), and $b_x$ and $b_y$ (4.26\,\AA\ and 2.46\,\AA, respectively) are the graphene lattice spacings along $x$ and $y$.
We have estimated $E\simeq36.8$\,GPa (compared with a bulk value of $78$\,GPa), and
$\tau_{max}\simeq5.92$\,meV/\AA$^3$, which, assuming the bulk value of $\nu=0.42$, yield 
$b_{core}\approx6.3$\,nm\,$\approx14$\,$a_x$.
\vfill
\newpage
\section{Bond-length distributions in the finite 2D FK model}\label{secS3:bond-distribution_FK}
\begin{figure}[!h]\label{figS2:bondFK}
  \centering
  \includegraphics[width=0.9\columnwidth]{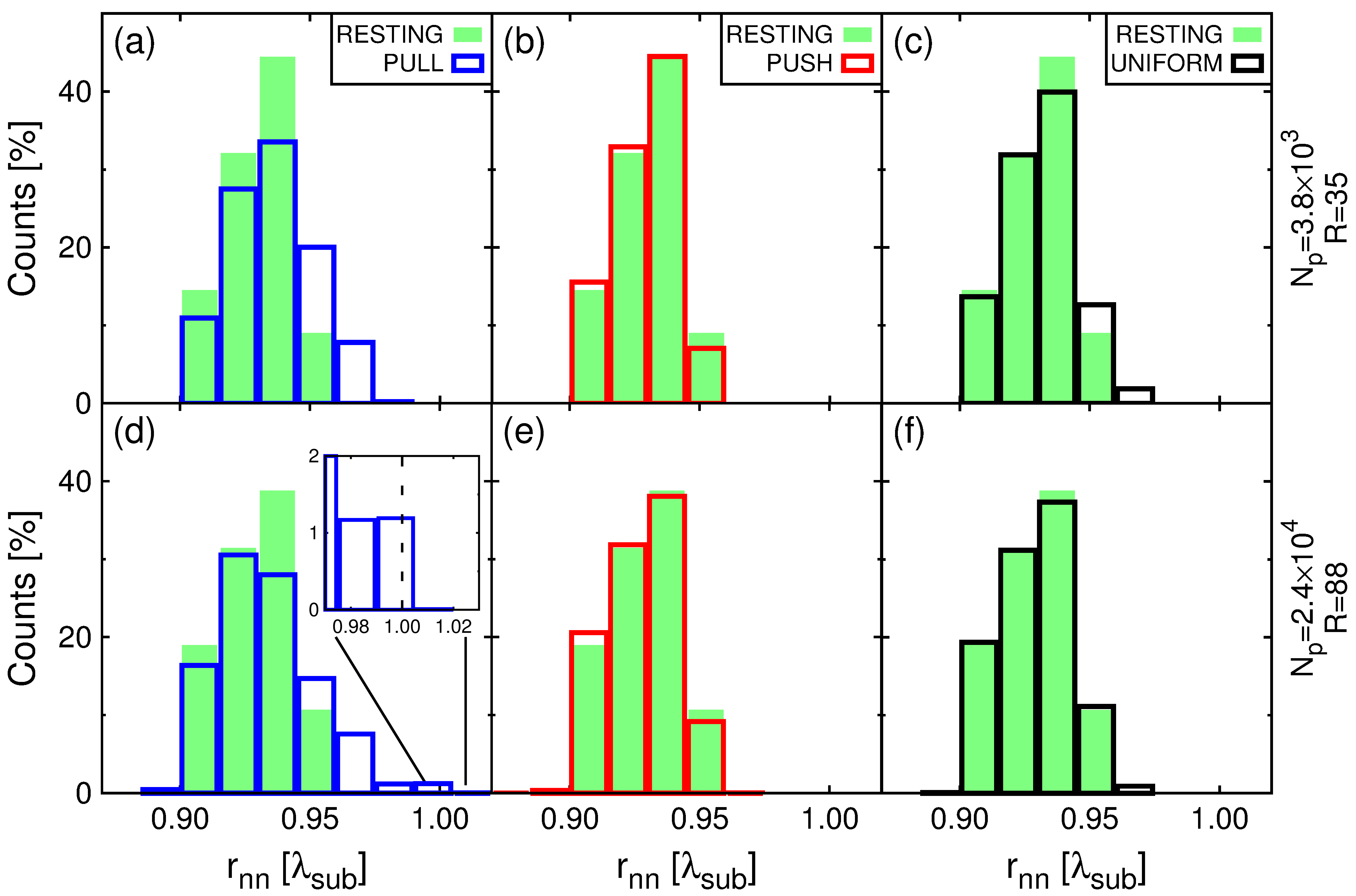}
  \caption{\small Distributions of the nearest-neighbor distances $r_{nn}$ (in units of the substrate periodicity
           $\lambda_{sub}$), measured in the circular islands of size $N_p=3.8\cdot10^3$ (panels (a),(b),(c))
           and $N_p=2.4\cdot10^4$ (panels (d),(e),(f)).
           From left to right, empty boxes are histograms obtained during pulling (blue), pushing (red), and
           uniform driving (black) at an applied external force $F_{ext}\lesssim F_s$, just before depinning.
           For comparison, the histograms (filled green boxes) corresponding to the
           relaxed configurations at rest ($F_{ext}=0$) are also reported. All simulations
           were performed adopting the following set of parameters of our 2D FK model: $a=1$, $k_1=10$, $k_2=5$,                  $\lambda_{sub}\approx(1+\sqrt{5})/3$, $U_0=0.075$.
          }
\end{figure}

Figure \ref{figS2:bondFK} reports the distributions of the nearest-neighbor distances $r_{nn}$
measured in the circular islands of sizes $N_p=3.8\cdot10^3$ (top panels) and $N_p=2.4\cdot10^4$
(bottom panels), respectively below and above the critical value $N_p^*\simeq 1\cdot10^4$
for the nucleation of a commensurate
dislocation. In both cases, the pulling protocol induces relatively large bond elongations
(see panels (a),(d)), which, in the larger island, reach local commensuration ($r_{nn}/\lambda_{sub}=1$, see
inset in panel (d)).
Compared to the bond distributions at rest, pushing is responsible for small compressions
(see panels (b),(e)), while a uniform driving results in mild elongations (see panels (c),(f)).
\vfill
\vfill
\newpage
\section{Deformations under sliding in the finite 2D FK model}\label{secS4:FK_sliding}
\begin{figure}[!h]\label{figS3:delta_loc-sliding}
  \centering
  \includegraphics[width=0.9\columnwidth]{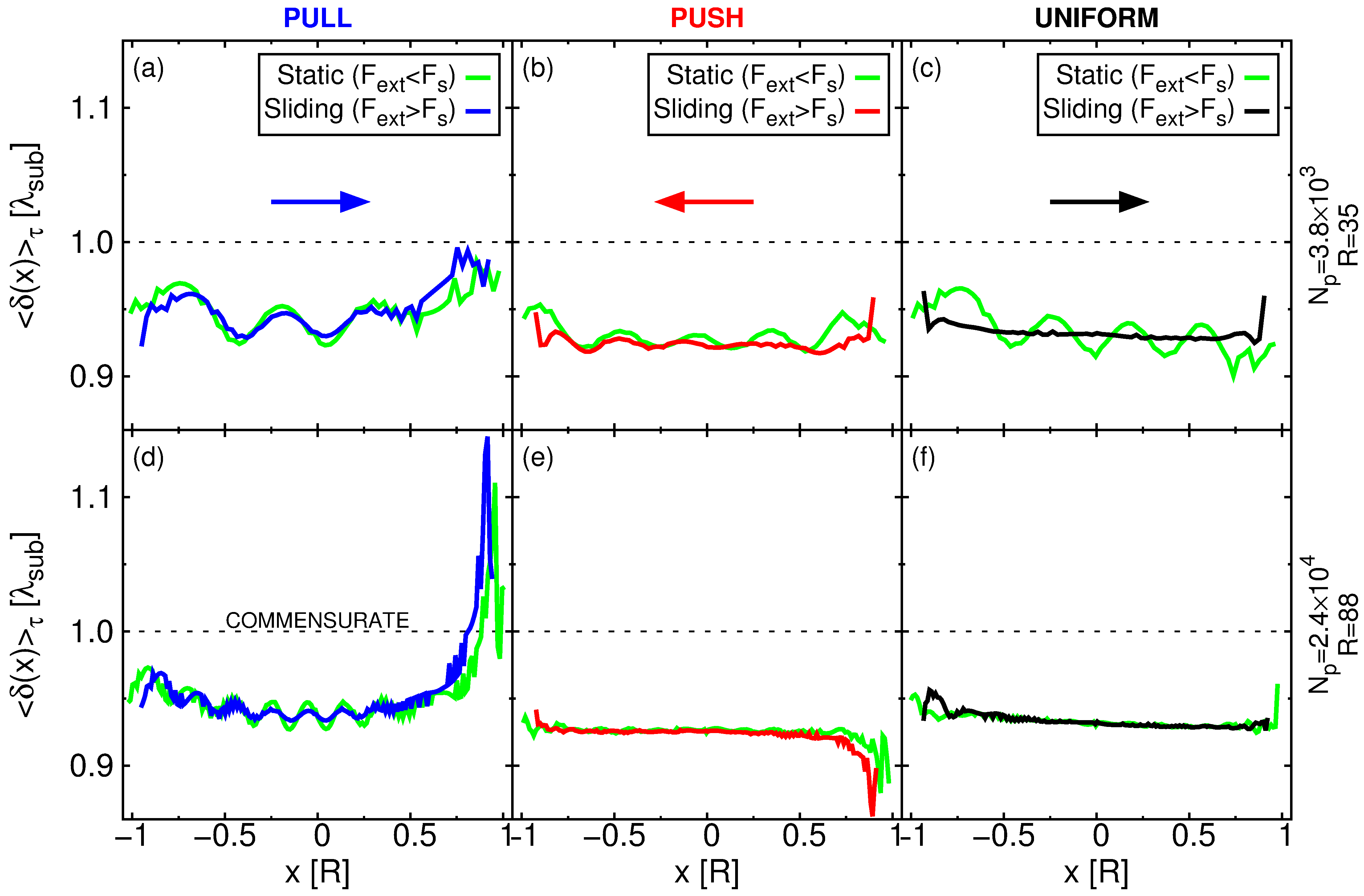}
  \caption{\small
          The averaged local mismatch $\delta(x)$ (see main text for defiinition) of two circular islands of
          the 2D FK model simulations is shown. A value of $\delta(x)=1$ indicates local commensuration
          to the substrate. Panels (a),(b),(c) and (d),(e),(f) are results for two sizes of $N_p=3.8\cdot 10^3$
          and $N_p=2.4\cdot 10^4$ particles, corresponding respectively to radii of $R=35$ and $R=88$.
          From left to right: blue, red, and black lines show the time average $\langle\delta(x)\rangle_{\tau}$
          computed under steady sliding for the pulling, pushing, and uniform protocol.
          For comparison, in each panel we also report the static values of $\delta(x)$ (green lines)
          measured before the onset of motion, at
          an applied external force $F_{ext}\lesssim F_s$, just below the depinning threshold. Arrows indicate the
          direction of $F_{ext}$ in the three protocols.
          All simulations were performed adopting the following set of parameters of our 2D FK model:
          $a=1$, $k_1=10$, $k_2=5$, $\lambda_{sub}\approx(1+\sqrt{5})/3$. For the sliding simulations we adopted
          a viscous damping $\eta=0.5$, which is necessary in order to reach steady state.
          }
\end{figure}

Figure~\ref{figS3:delta_loc-sliding} reports the time average $\langle \delta(x) \rangle_{\tau}$ of the local
mismatch measured under steady-sliding of the circular islands of sizes $N_p=3.8\cdot10^3$ (top panels)
and $N_p=2.4\cdot10^4$ (bottom panels). Compared to the static configurations before the onset of motion
(green lines in Fig.~\ref{figS3:delta_loc-sliding}), we observe a general enhancement
of the local lattice distortions. A localized commensurate dislocation appears only under edge-pulling of the
larger island. During sliding, the dislocation moves towards the bulk, but remains
close to the leading edge (see panel (d)). A similar behavior was observed up to the largest size investigated.
\vfill
\vfill
\newpage
\section{Scaling exponents in the atomistic simulations}\label{secS5:ScalingGold}
\begin{figure}[!h]\label{figS4:scaling-gold}
  \centering
  \includegraphics[width=0.8\columnwidth]{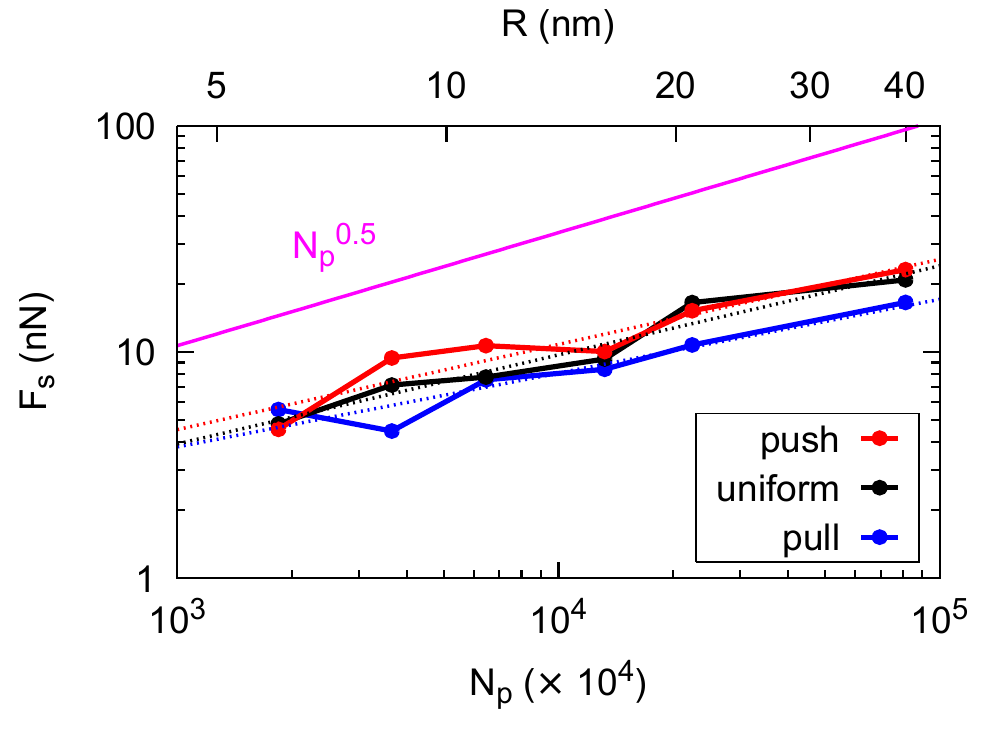}
  \caption{\small
           Log-Log plot showing the size scaling of the static friction force $F_s$ as a function of the number of 
           particles $N_p$ (bottom $x$-axis) and of the radius $R$ (upper $x$-axis) of model gold islands sliding
           over graphene. Blue, black, and red lines are results obtained during the pulling, uniform, and pushing 
           protocols, respectively. Dotted lines are linear regressions of each data set.
           The magenta line shows the square root scaling law $F_s\propto N_p^{1/2}$, 
           corresponding to the maximum value $\gamma_{max}=1/2$ of the scaling exponent, expected for 
           incommensurate superlubric contacts.
          }
\end{figure}

Figure~\ref{figS4:scaling-gold} is a Log-Log plot of the size scaling of static friction in the
gold-graphene system. A linear fit of the data yielded the following scaling exponents: 
$\gamma_{gold}^{pull}=0.33\pm0.06$, $\gamma_{gold}^{uniform}=0.39\pm0.05$, and $\gamma_{gold}^{push}=0.38\pm0.07$. 
The limited amount of data do not allow to obtain accurate values of $\gamma$, that are affected by 
relative errors of $\sim10-20\%$.
\vfill
\vfill
\newpage